\documentclass[fp,twocolumn]{jpsj3}
\usepackage{txfonts}
\usepackage{color,ulem}

\title{
Time-of-Flight Elastic and Inelastic Neutron Scattering Studies on the Localized $4d$ Electron Layered Perovskite La$_5$Mo$_4$O$_{16}$
}

\author{
Kazuki~Iida$^1$\thanks{k\_iida@cross.or.jp}, 
Ryoichi~Kajimoto$^2$, 
Yusuke~Mizuno$^3$, 
Kazuya~Kamazawa$^1$, 
Yasuhiro~Inamura$^2$, 
Akinori~Hoshikawa$^4$, 
Yukihiko~Yoshida$^4$, 
Takeshi~Matsukawa$^4$, 
Toru~Ishigaki$^4$, 
Yukihiko~Kawamura$^1$, 
Soshi~Ibuka$^5$, 
Tetsuya~Yokoo$^5$, 
Shinichi~Itoh$^5$, and 
Takuro~Katsufuji$^{3,6}$
}

\inst{
$^1$Neutron Science and Technology Center, Comprehensive Research Organization for Science and Society (CROSS), Tokai, Ibaraki 319-1106, Japan\\
$^2$J-PARC Center, Japan Atomic Energy Agency (JAEA), Tokai, Ibaraki 319-1195, Japan\\
$^3$Department of Physics, Waseda University, Shinjuku, Tokyo 169-8555, Japan\\
$^4$Frontier Research Center for Applied Atomic Sciences, Ibaraki University, Tokai, Ibaraki 319-1106, Japan\\
$^5$Neutron Science Division, Institute of Materials Structure Science, High Energy Accelerator Research Organization (KEK), Tsukuba 305-0801, Japan\\
$^6$Kagami Memorial Research Institute for Materials Science and Technology, Waseda University, Tokyo 169-0051, Japan\\
}

\abst{
The magnetic structure and spin-wave excitations in the quasi-square-lattice layered perovskite compound La$_5$Mo$_4$O$_{16}$ were studied by a combination of neutron diffraction and inelastic neutron scattering techniques using polycrystalline sample.
Neutron powder diffraction refinement revealed that the magnetic structure is ferrimagnetic in the $ab$ plane with antiferromagnetic stacking along the $c$ axis where the magnetic propagation vector is $\mathbf{k}=\left(0,0,\frac{1}{2}\right)$.
The ordered magnetic moments are estimated to be $0.54(2)\mu_\text{B}$ for Mo$^{5+}$ ($4d^1$) ions and $1.07(3)\mu_\text{B}$ for Mo$^{4+}$ ($4d^2$) ions at 4~K, which are about half of the expected values.
The inelastic neutron scattering results display strong easy-axis magnetic anisotropy along the $c$ axis due to the spin-orbit interaction in Mo ions evidenced by the spin gap at the magnetic zone center.
The model Hamiltonian consisting of in-plane anisotropic exchange interactions, the interlayer exchange interaction, and easy-axis single-ion anisotropy can explain our inelastic neutron scattering data well.
Strong Ising-like anisotropy and weak interlayer coupling compared with the intralayer exchange interaction can explain both the high-temperature magnetoresistance and long-time magnetization decay recently observed in La$_5$Mo$_4$O$_{16}$.
}

\begin{document}
\maketitle

\section{Introduction}

Recently, $4d$ localized magnetic systems have attracted much attention in condensed matter physics because they show abundant magnetic phenomena due to the intermediate scale of the spin-orbit interaction (SOI).
For instance, the exotic spin liquid ground state realized in the face-centered-cubic double perovskite Ba$_2$YMoO$_6$~\cite{Ba2YMoO6_1} and the Kitaev quantum spin liquid in the two-dimensional honeycomb lattice $\alpha$-RuCl$_3$~\cite{RuCl3_1} have been investigated intensively in the framework of their relatively large SOI.
In both systems, the considerable SOI plays an important role in the magnetic phenomena.
It is important to investigate the magnetism with $4d$ localized magnetic moments in other lattices such as a square lattice, which can also be frustrated in the presence of the next-nearest-neighbor antiferromagnetic interaction.
Therefore, the magnetic properties in the quasi-two-dimensional square lattice system La$_5$Mo$_4$O$_{16}$ with localized $4d$ electrons are investigated in this regard.

La$_5$Mo$_4$O$_{16}$ is a layered perovskite compound with monoclinic symmetry ($C2/m$, No.~12), and there are three inequivalent Mo sites (Mo1, Mo2, and Mo3 as shown later in Fig.~\ref{Fig:MagneticStructure}).~\cite{La5Mo4O16_McCarroll,La5Mo4O16_Ledesert}
Corner-sharing MoO$_6$ octahedra consisting of Mo1 and Mo2 form a quasi-square checkerboard lattice.
Meanwhile, two Mo3 form a Mo$_2$O$_{10}$ pillar, which is located between the perovskite layers and connects Mo2 octahedra.
The valences of the Mo1, Mo2, and Mo3 sites are 5+, 4+, and 4+, respectively~\cite{La5Mo4O16_Ledesert}, and the Mo1 and Mo2 sites have spins $S=1/2$ ($4d^1$) and $S=1$ ($4d^2$).~\cite{La5Mo4O16_Lofland}
On the other hand, the molecular orbital in the edge-sharing bioctahedral Mo$_2$O$_{10}$ is in the low spin state.~\cite{La5Mo4O16_Ramanujachary1}
Since Mo$_2$O$_{10}$ pillars are nonmagnetic and the distance between the layers is large, the interlayer interaction is expected to be very weak.
Susceptibility measurement on polycrystalline La$_5$Mo$_4$O$_{16}$ shows an antiferromagnetic phase transition at $T_\text{N}=190$~K.~\cite{La5Mo4O16_Ramanujachary1,La5Mo4O16_Ramanujachary2}
In addition, single-crystal magnetization measurements on La$_5$Mo$_4$O$_{16}$ exhibited a metamagnetic transition from the antiferromagnetic to ferrimagnetic state above 0.5~T.~\cite{La5Mo4O16_Kobayashi}
This ferrimagnetic phase is only observed when an external magnetic field is applied along the $c$ axis, suggesting easy-axis magnetic anisotropy.~\cite{La5Mo4O16_Kobayashi,La5Mo4O16_Lofland,La5Mo4O16_Konishi}

Recently, intriguing magnetic phenomena in La$_5$Mo$_4$O$_{16}$ were reported.
High-temperature magnetoresistance up to 180~K~\cite{La5Mo4O16_Kobayashi} was observed as in the Ising-like layered compound SrCo$_6$O$_{11}$.~\cite{SrCo6O11}
In addition, long-time magnetization decay from the field-induced ferrimagnetic to zero-field antiferromagnetic state was also observed within the temperature range of $35<T<55$~K.~\cite{La5Mo4O16_Konishi}
Both the high-temperature magnetoresistance and long-time magnetization decay were explained by uniaxial anisotropy and weak interlayer coupling compared with intralayer coupling.
Therefore, easy-axis anisotropy and quasi-two dimensionality are the key to understanding the fascinating magnetism in La$_5$Mo$_4$O$_{16}$, which has never been investigated by spectroscopic techniques.
Furthermore, the magnetic structures in La$_5$Mo$_4$O$_{16}$ have not been investigated by neutron diffraction because neither large single crystals nor large amounts of polycrystalline samples have been synthesized.

In this paper, we report elastic and inelastic neutron scattering results on polycrystalline La$_5$Mo$_4$O$_{16}$ without an external magnetic field to determine the zero-field magnetic structure and model Hamiltonian.
The obtained magnetic structure is ferrimagnetic in the $ab$ plane but antiferromagnetic along the $c$ axis with the magnetic propagation vector $\mathbf{k}=\left(0,0,\frac{1}{2}\right)$.
Reduced magnetic moments of $0.54(2)\mu_\text{B}$ and $1.07(3)\mu_\text{B}$ for Mo1 and Mo2 at 4~K were observed, respectively.
In the inelastic neutron scattering (INS) spectra, a clear spin gap of $\sim9$~meV was observed at the magnetic zone center, suggesting strong magnetic anisotropy.
A model Hamiltonian consisting of in-plane anisotropic exchange interactions, the interlayer exchange interaction, and easy-axis single-ion anisotropy was determined by comparison of linear spin wave (LSW) calculations with INS results.
We attribute the easy-axis magnetic anisotropy to the sizable SOI in Mo ions, whereas the reduced magnetic moments are attributed to fluctuations induced by quantum effects, two dimensionality, and $J_1$--$J_2$ bond frustration.

\section{Experimental Methods}
Polycrystalline La$_5$Mo$_4$O$_{16}$ for zero-field elastic and inelastic neutron scattering measurements was prepared by solid state reactions.
Neutron powder diffraction (NPD) patterns at $T=4$, 40, 100, and 220~K were taken at the time-of-flight (TOF) neutron diffractometer iMATERIA~\cite{iMATERIA} installed at BL20 in the Materials and Life Science Experimental Facility (MLF), Japan Proton Accelerator Research Complex (J-PARC).
Approximately 3.7~g of powder was packed into a vanadium cell.
The TOF neutron diffraction results presented in this paper were obtained by the $90^\circ$ detector bank.
Refinements of crystal and magnetic structures were carried out by the Rietveld method using the GSAS program.~\cite{Refinement}
INS measurements were performed at the TOF neutron spectrometers 4SEASONS~\cite{4SEASONS_1,4SEASONS_2} and HRC~\cite{HRC_1,HRC_2} installed at BL01 and BL12, respectively, in MLF, J-PARC.
Roughly 11 and 50~g of powder were used for the 4SEASONS and HRC measurements, and the measurements were performed at $T=4$, 103, and 224~K.
INS data were visualized by software suites Utsusemi.~\cite{4SEASONS_3}
Neutron incident energy ($E_\text{i}$) of 15.3~meV with a Fermi chopper rotating at a frequency ($f_\text{Fermi}$) of 250~Hz was used at 4SEASONS, whereas incident neutrons with $E_\text{i}=102.4$~meV and $f_\text{Fermi}=400$~Hz were used at HRC.
The estimated energy resolutions at the elastic channel for $E_\text{i}=15.3$ and 102.4~meV were 0.66 and 3.87~meV, respectively.~\cite{HRC_1,4SEASONS_4}
The time-independent background was subtracted from the data at 4SEASONS.~\cite{4SEASONS_5}
In the present study, we used the magnetic form factor of Mo$^+$ in Ref.~\citen{MFfactor} for both Mo$^{5+}$ and Mo$^{4+}$.
All indexes in this paper are expressed by the $C2/m$ space group.

\section{Results and Discussion}
\begin{figure}[t]
\includegraphics[width=8.54cm]{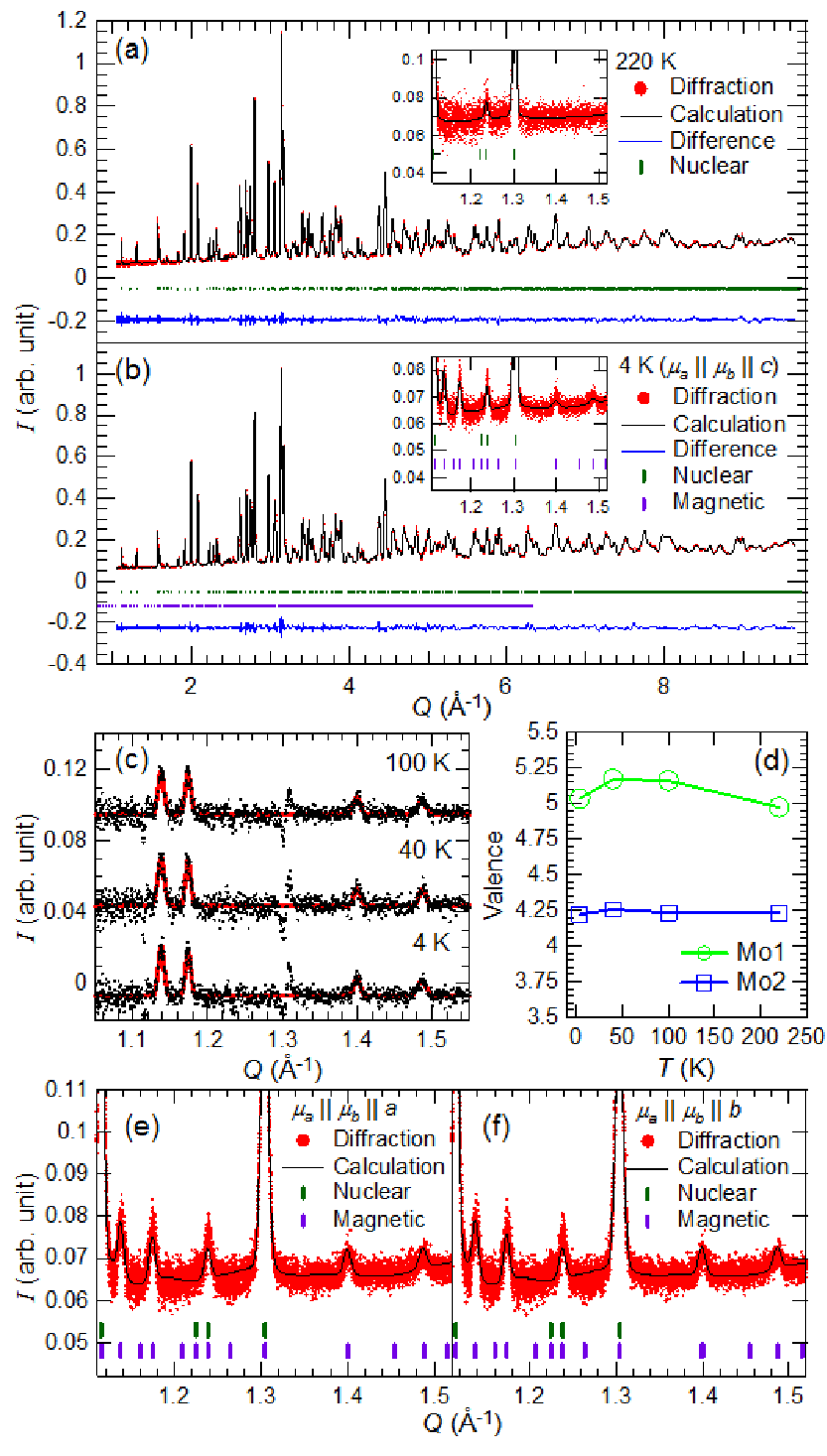}
\caption{\label{Fig:Diffraction}
(Color online)
NPD patterns (red) in La$_5$Mo$_4$O$_{16}$ at (a) $T=220$ and (b) 4~K measured by iMATERIA.
Calculated (black) and difference (blue) patterns are also shown.
Upper (green) and lower (purple) ticks represent positions of nuclear and magnetic Bragg peaks, respectively.
The weighted profile residuals at 4, 40 (not shown), 100 (not shown), and 220 K are $R_\text{wp}=3.86$\%, $3.91$\%, $3.85$\%, and $3.77$\%, respectively.
Refined structural parameters at 4~K are summarized in Table.~\ref{table}, and the obtained magnetic structure is shown in Fig.~\ref{Fig:MagneticStructure}.
The insets in panels (a) and (b) show magnified views.
(c) NPD patterns at 4, 40, and 100~K, where the diffraction pattern at 220~K was subtracted as background.
The solid lines are Gaussian fitting results.
Data at 40 and 100~K are shifted vertically for clarity.
(d) Temperature dependences of valences at the Mo1 and Mo2 sites determined by a bond valence sum method~\cite{BVS} using the refined crystal parameters.
Magnified views of Rietveld refinement results at 4~K assuming the magnetic structures (e) $\mathbf{\mu}_1\parallel\mathbf{\mu}_2\parallel a$ and (f) $\mathbf{\mu}_1\parallel\mathbf{\mu}_2\parallel b$.
The weighted profile residuals are (e) $R_\text{wp}=3.86$\% and (f) $3.91$\%.
The obtained magnetic moments are (e) $\mu_1=0.63(2)\mu_\text{B}$ and $\mu_2=1.26(4)\mu_\text{B}$ and (f) $\mu_1=0.64(2)\mu_\text{B}$ and $\mu_2=1.29(4)\mu_\text{B}$.
}
\end{figure}

A NPD pattern in La$_5$Mo$_4$O$_{16}$ at $T=220$~K ($>T_\text{N}$) together with a Rietveld refinement result is shown in Fig.~\ref{Fig:Diffraction}(a).
We started with the crystal information reported in the previous single-crystal X-ray diffraction study~\cite{La5Mo4O16_Ledesert} and then refined it.
Reasonable agreement can be seen between the observed and calculated NPD patterns, suggesting the goodness of the refined structural parameters.

Figure~\ref{Fig:Diffraction}(c) shows NPD patterns at 4, 40, and 100~K ($<T_\text{N}$), in which the diffraction pattern at 220~K has been subtracted as background so that only magnetic Bragg peaks are visible.
Upon decreasing the temperature below $T_\text{N}$, magnetic Bragg peaks appear at momentum transfers ($Q$s) of 1.138(2), 1.174(2), 1.400(6), and 1.486(8)~$\text{\AA}^{-1}$.
The peak positions correspond to $\left(\overline{1}1\frac{1}{2}\right)$, $\left(11\frac{1}{2}\right)$, $\left(\overline{1}1\frac{3}{2}\right)$, and $\left(11\frac{3}{2}\right)$, and the magnetic propagation vector can be indexed by $\mathbf{k}=\left(0,0,\frac{1}{2}\right)$.
The difference between the spectra at 4, 40, and 100~K is very small, indicating that the magnetic structures at 4, 40, and 100~K are the same.
Although single-crystal magnetization measurements under the external magnetic field along the $c$ axis show an additional ferrimagnetic phase below 70~K,~\cite{La5Mo4O16_Kobayashi} no ferrimagnetic signature was observed in the zero-field NPD patterns at 4 and 40~K [Fig.~\ref{Fig:Diffraction}(c)].
These results indicate that La$_5$Mo$_4$O$_{16}$ has a strong easy-axis magnetic anisotropy and that the direction of the external magnetic field is important for observing the ferrimagnetic state as well as the long-time magnetization decay.~\cite{La5Mo4O16_Konishi}

\begin{figure}[t]
\includegraphics[width=7.3cm]{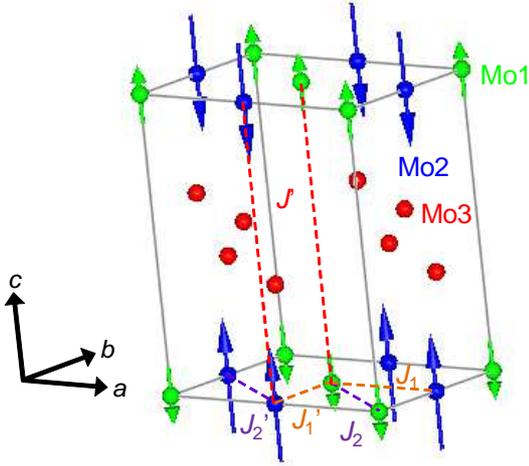}
\caption{\label{Fig:MagneticStructure}
(Color online)
Zero-field magnetic structure of La$_5$Mo$_4$O$_{16}$ with the magnetic propagation vector $\mathbf{k}=\left(0,0,\frac{1}{2}\right)$ drawn by the software suite FullProf.~\cite{FullProf}
Spheres represent the Mo1 (green), Mo2 (blue), and Mo3 (red) sites, respectively, while dashed lines represent exchange interactions.
Small (green) and large (blue) arrows represent size of the magnetic moments [$\mu_1=0.54(2)\mu_\text{B}$ and $\mu_2=1.07(3)\mu_\text{B}$] obtained by the Rietveld refinement at 4~K.
Solid lines show the chemical unit cell.
}
\end{figure}

A magnetic structure, where Mo1 and Mo2 show collinear antiferromagnetic coupling within the layers, that is in-plane ferrimagnetic and the layers are also antiferromagnetically coupled along the $c$ axis, as described in Fig.~\ref{Fig:MagneticStructure}, was used to analyze the NPD patterns below $T_\text{N}$.
We considered the same magnetic structure as those of La$_5$Mo$_{2.75}$V$_{1.25}$O$_{16}$,~\cite{La5VMo4O16_Ramezanipour} La$_5$Os$_3$MnO$_{16}$,~\cite{La5Os3MnO16_1} and La$_5$Re$_3$CoO$_{16}$~\cite{La5Re3MO16_2} because these compounds have a similar crystal structure and magnetic propagation vector to La$_5$Mo$_4$O$_{16}$.
Since the magnetic structure factors of all the observed magnetic reflections [$\left(\overline{1}1\frac{1}{2}\right)$, $\left(11\frac{1}{2}\right)$, $\left(\overline{1}1\frac{3}{2}\right)$, and $\left(11\frac{3}{2}\right)$ as shown in Fig.~\ref{Fig:Diffraction}(c)] are proportional to $\mu_1+\mu_2$, where $\mu_1$ and $\mu_2$ are the ordered magnetic moments at Mo1 and Mo2, respectively, we assume the constraint on the magnetic moments $2\mu_1=\mu_2$.
The best fit, as seen in the difference plot in Fig.~\ref{Fig:Diffraction}(b), was obtained when the magnetic moments are parallel to the $c$ axis with $\mu_1=0.54(2)\mu_\text{B}$ and $\mu_2=1.07(3)\mu_\text{B}$.
The refined structural parameters at 4~K are summarized in Table.~\ref{table}.
A previously reported single-crystal magnetization measurement found that the field-induced ferrimagnetic state in La$_5$Mo$_4$O$_{16}$ has a magnetic moment of $\sim0.5\mu_\text{B}$/f.u. where Mo1 and Mo2 sites are antiferromagnetically coupled (i.e. $\mu_2-\mu_1$).~\cite{La5Mo4O16_Lofland,La5Mo4O16_Kobayashi}
The good agreement between the neutron diffraction and single-crystal bulk magnetization supports the validity of our magnetic structure.
Although a magnetic structure with $\mathbf{\mu}_1\parallel\mathbf{\mu}_2\parallel a$ or $\mathbf{\mu}_1\parallel\mathbf{\mu}_2\parallel b$, which is also a possible magnetic structure of La$_5$Mo$_4$O$_{16}$ with $\mathbf{k}=\left(0,0,\frac{1}{2}\right)$, can give a similar fitting result [Figs.~\ref{Fig:Diffraction}(e) and \ref{Fig:Diffraction}(f)], such magnetic structures are ruled out by the single-crystal magnetization measurements, which reported the easy-axis anisotropy along the $c$ axis.~\cite{La5Mo4O16_Kobayashi,La5Mo4O16_Lofland}

\begin{table}[b]
\caption{
Structural parameters of La$_5$Mo$_4$O$_{16}$ at 4~K obtained from our Rietveld refinement shown in Fig.~\ref{Fig:Diffraction}(b).
The refined lattice constants are $a=7.96252(9)$~$\text{\AA}$, $b=8.01163(7)$~$\text{\AA}$, $c=10.2996(6)$~$\text{\AA}$, and $\beta=94.968(1)^{\circ}$.
The occupancies at the O6 and O7 sites are 0.5.~\cite{La5Mo4O16_Ledesert}
}
\label{table}
\begin{center}
\begin{tabular}{lllll}
\hline
\hline
atom & site & $x$ & $y$ & $z$\\
\hline
La1 & $2c$ & 0         & 0         & 0.5      \\
La2 & $8j$ & 0.2715(1) & 0.2463(1) & 0.2138(1)\\
Mo1 & $2a$ & 0         & 0         & 0        \\
Mo2 & $2b$ & 0.5       & 0         & 0        \\
Mo3 & $4i$ & 0.5612(2) & 0         & 0.3994(1)\\
O1  & $8j$ & 0.0356(1) & 0.2572(2) & 0.3572(1)\\
O2  & $4i$ & 0.2809(2) & 0.5       & 0.3283(2)\\
O3  & $4i$ & 0.3060(2) & 0         & 0.4297(2)\\
O4  & $4i$ & 0.4608(2) & 0         & 0.1872(2)\\
O5  & $4i$ & 0.0778(2) & 0         & 0.1738(2)\\
O6  & $8j$ & 0.0440(2) & 0.2388(3) & 0.9941(3)\\
O7  & $8j$ & 0.2487(3) & 0.9709(1) & 0.9594(2)\\
\hline
\hline
\end{tabular}
\end{center}
\end{table}

Considering $S=1/2$ and 1 for Mo1 and Mo2 and $g=2$, the obtained magnetic moments at 4~K are about half of the expected values ($1\mu_\text{B}$ and $2\mu_\text{B}$).
Neutron diffraction measurements on isostructural V-doped La$_5$Mo$_{2.75}$V$_{1.25}$O$_{16}$ also found the reduced magnetic moments of Mo ions.~\cite{La5VMo4O16_Ramezanipour}
Thus, the SOI in such layered Mo compounds does not give rise to the total momentum number $\mathbf{J}=\mathbf{S}+\mathbf{L}$, which results in large magnetic moments as in $5d$ compounds.
On the other hand, our NPD refinement and the field-induced ferrimagnetic structure being sensitive to the external magnetic field direction~\cite{La5Mo4O16_Kobayashi} suggest that the SOI in La$_5$Mo$_4$O$_{16}$ gives the easy-axis magnetic anisotropy.
The origin of the reduced magnetic moments will be discussed later by combining elastic and inelastic neutron scattering results.

\begin{figure}[t]
\includegraphics[width=8.54cm]{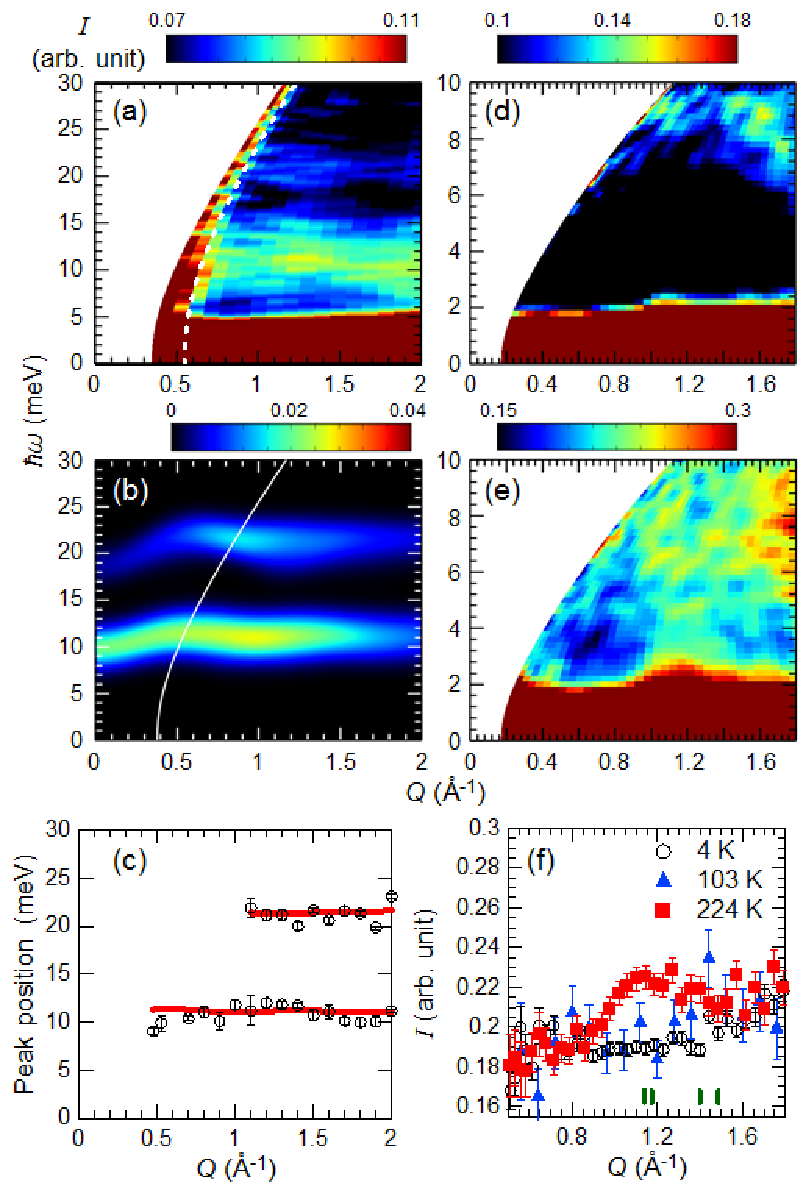}
\caption{\label{Fig:Map}
(Color online)
(a) $I(Q,\hbar\omega)$ map in La$_5$Mo$_4$O$_{16}$ at $T=4$~K measured by HRC with $E_\text{i}=102.4$~meV.
The white dashed line represents the scan trajectory along $2\theta=4.5^\circ$.
(b) Calculated $I(Q,\hbar\omega)$ map at 4~K using Eqs.~(\ref{Eq:Hamiltonian})$-$(\ref{Eq:Parameters}).
Instrumental energy resolution at HRC assuming $E_\text{i}=102.4$~meV was convoluted.
The white line shows the detector coverage boundary.
(c) $Q$ dependences of peak positions of two magnetic excitations in La$_5$Mo$_4$O$_{16}$ determined by Gaussian fitting at each $Q$ position.
The solid lines are calculated peak positions using Eqs.~(\ref{Eq:Hamiltonian})$-$(\ref{Eq:Parameters}).
$I(Q,\hbar\omega)$ maps measured by 4SEASONS with $E_\text{i}=15.3$~meV at (d) 4 and (e) 224~K.
(f) $Q$ dependences of neutron scattering intensities at 4, 103, and 224~K with $E_\text{i}=15.3$~meV.
$\hbar\omega$ was integrated in $[3,6.5]$~meV at each temperature.
Data at 4 and 103 K are vertically shifted by 0.12 and 0.095, respectively, to match the background intensities at $Q<0.8$ and $Q>1.6$~$\text{\AA}^{-1}$.
Vertical bars represent the magnetic Bragg positions observed in the NPD patterns in Figs.~\ref{Fig:Diffraction}(b) and \ref{Fig:Diffraction}(c).
}
\end{figure}

The INS intensity ($I$) map in La$_5$Mo$_4$O$_{16}$ at $T=4$~K as a function of $Q$ and energy transfer ($\hbar\omega$) is shown in Fig.~\ref{Fig:Map}(a).
A broad-energy excitation centered around 10~meV was observed.
Another excitation also exists at 20~meV.
Both excitations have magnetic origin because neutron scattering intensities at both 10 and 20~meV decrease monotonically with increasing $Q$ as described later.
It should be noted that the strong signal at a small scattering angle ($2\theta<4.5^\circ$) is due to contamination from the direct neutron beam.
Since magnetic Bragg peaks are located at $Q=1.14$, 1.17, 1.40, and 1.49~$\text{\AA}^{-1}$, as plotted in Fig.~\ref{Fig:Diffraction}(c), Goldstone modes of spin-wave excitations are expected to evolve in the vicinity of the magnetic zone centers below $T_\text{N}$.
However, as shown in Fig.~\ref{Fig:Map}(d), the $I(Q,\hbar\omega)$ map using lower $E_\text{i}$ with better energy resolution clearly shows the absence of Goldstone modes in $1\le Q\le1.5$~$\text{\AA}^{-1}$, and a spin gap of $\sim9$~meV is observed.
On the other hand, as shown in Fig.~\ref{Fig:Map}(e), the $I(Q,\hbar\omega)$ map at 224~K ($>T_\text{N}$) shows a gapless quasielastic excitation centered around the magnetic zone centers, which is a common feature of cooperative paramagnets.~\cite{SHL}
We cut $I(Q,\hbar\omega)$ maps at 4, 103, and 224~K along the $Q$ direction by integrating in $\hbar\omega=[3,6.5]$~meV, and the obtained $I(Q)$ cuts are shown in Fig.~\ref{Fig:Map}(f).
Neutron scattering intensities at 4 and 103~K ($<T_\text{N}$) are comparable with the background, while the intensity at 224~K ($>T_\text{N}$) is clearly enhanced around the magnetic zone centers.
Therefore, the magnetic excitation at 10~meV is assigned to a gapped spin-wave excitation in the long-range magnetic ordered state.
The spin gap at the magnetic zone center below $T_\text{N}$ exhibits strong Ising-like magnetic anisotropy originating from the SOI as expected from the single-crystal bulk magnetization measurements.~\cite{La5Mo4O16_Kobayashi,La5Mo4O16_Lofland}
Although there is no clear evidence that the magnetic excitation at 20~meV is a spin-wave excitation, the 20~meV mode can be well reproduced by our LSW calculation.

\begin{figure}[t]
\includegraphics[width=8.54cm]{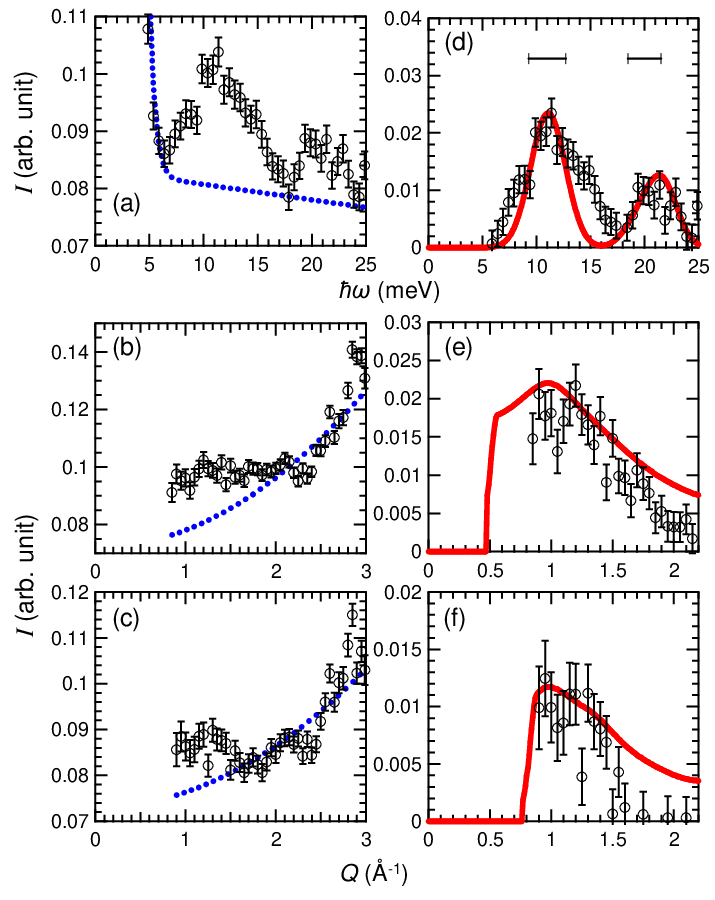}
\caption{\label{Fig:Cuts}
(Color online)
(a) $\hbar\omega$ dependence of neutron scattering intensities in La$_5$Mo$_4$O$_{16}$ at $T=4$~K.
$Q$ was integrated in $[1.0,1.4]~\text{\AA}^{-1}$.
$Q$ dependences of neutron scattering intensities at 4~K in (b) $\hbar\omega=[9,12]$ and (c) $[19,22]$~meV.
The experimental results in panels (a)--(c) were measured by HRC with $E_\text{i}=102.4$~meV.
The dashed lines in panels (a)--(c) are estimated background spectra described in the main text.
(d)--(f) Background subtracted $\hbar\omega$ and $Q$ dependences of neutron scattering intensities at 4~K.
The solid lines in panels (d)--(f) are calculated LSW intensities at 4~K using Eqs.~(\ref{Eq:Hamiltonian})$-$(\ref{Eq:Parameters}).
The horizontal bars in panel (d) are instrumental energy resolutions at $\hbar\omega=11$ and 20~meV.~\cite{HRC_1,4SEASONS_4}
}
\end{figure}

In order to better quantify the spin-wave excitations in La$_5$Mo$_4$O$_{16}$, the $\hbar\omega$ and $Q$ dependences of the neutron scattering intensities at 4~K are plotted in Figs.~\ref{Fig:Cuts}(a)--\ref{Fig:Cuts}(c).
Figure~\ref{Fig:Cuts}(a) shows the $\hbar\omega$ dependence of the neutron scattering intensities at $Q=[1.0,1.4]$~$\text{\AA}^{-1}$.
To isolate the magnetic scattering, the linear background and the incoherent scattering centered at the elastic channel were fit to the raw energy spectrum and then subtracted.
The background-subtracted energy spectrum is plotted in Fig.~\ref{Fig:Cuts}(d).
There are two excitation peaks at $\hbar\omega\sim11$ and 20~meV.
The instrumental energy resolutions at $\hbar\omega=11$ and 20~meV are described by horizontal bars.
The observed spin-wave excitations are broader than the instrumental energy resolutions, suggesting that various magnetic interactions exist.
Figures~\ref{Fig:Cuts}(b) and \ref{Fig:Cuts}(c) show the $Q$ dependences of the neutron scattering intensities at $\hbar\omega=[9,12]$ and $[19,22]$~meV, respectively.
To subtract the phonon contribution in the $Q$ dependences, a quadratic background was fitted to each raw $Q$ dependence in the range of $2.2\le Q\le 3.0$~$\text{\AA}^{-1}$ [dotted lines in Figs.~\ref{Fig:Cuts}(b) and \ref{Fig:Cuts}(c)].
The background-subtracted $Q$ dependences are plotted in Figs.~\ref{Fig:Cuts}(e) and \ref{Fig:Cuts}(f).
The neutron scattering intensities clearly decrease at high $Q$ in both $Q$ dependences, as expected for magnetic scattering.
To obtain the background in each spectrum precisely, we assumed the same background level at the equivalent $(Q,\hbar\omega)$ positions in the different spectra: $Q=1.2$~$\text{\AA}^{-1}$ and $\hbar\omega=10.5$~meV in Figs.~\ref{Fig:Cuts}(a) and \ref{Fig:Cuts}(b), and $Q=1.2$~$\text{\AA}^{-1}$ and $\hbar\omega=20.5$~meV in Figs.~\ref{Fig:Cuts}(a) and \ref{Fig:Cuts}(c).

As described above, all the features of the bulk magnetization and neutron scattering measurements suggest that La$_5$Mo$_4$O$_{16}$ has easy-axis (Ising) anisotropy due to the sizable SOI.
To explain our INS results, we now consider the following model Hamiltonian for La$_5$Mo$_4$O$_{16}$: 
\begin{eqnarray}
\mathcal{H}&=&J_1\sum_{i,j}\left[\delta_1\left(S_i^xT_j^x+S_i^yT_j^y\right)+S_i^zT_j^z\right]\nonumber\\
&+&J_1'\sum_{i,j}\left[\delta_1'\left(S_i^xT_j^x+S_i^yT_j^y\right)+S_i^zT_j^z\right]\nonumber\\
&+&J_2\sum_{i,j}\left[\delta_2\left(S_i^xS_j^x+S_i^yS_j^y\right)+S_i^zS_j^z\right]\nonumber\\
&+&J_2'\sum_{i,j}\left[\delta_2'\left(T_i^xT_j^x+T_i^yT_j^y\right)+T_i^zT_j^z\right]\label{Eq:Hamiltonian}\\
&+&J'\sum_{i,j}\mathbf{S}_i\cdot\mathbf{S}_j+J'\sum_{i,j}\mathbf{T}_i\cdot\mathbf{T}_j\nonumber\\
&+&D\sum_i\left(T_i^z\right)^2,\nonumber
\end{eqnarray}
where $J_n$, $\delta_n$, $J'$, and $D$ are the $n$th-nearest-neighbor exchange coupling (see Fig.~\ref{Fig:MagneticStructure}), the Ising parameter for the $n$th-nearest-neighbor exchange interaction, the interlayer exchange coupling, and easy-axis single-ion anisotropy, respectively.
$z$ is parallel to the $c$ axis, which is the easy axis.
$\mathbf{S}$ and $\mathbf{T}$ are the spin operators for spins 1/2 and 1, respectively.
Since single-crystal torque measurements reported no in-plane anisotropy,~\cite{La5Mo4O16_Lofland} we assume $J_1=J_1'$ and $\delta_1=\delta_1'$, $J_2=J_2'$ and $\delta_2=\delta_2'$.
Because $S=1/2$, the single-ion anisotropic term $\left(S_m^z\right)^2$ is not considered.
We diagonalized Eq.~(\ref{Eq:Hamiltonian}) by using a conventional semiclassical LSW theory.~\cite{LSW}

Neutron scattering intensities from magnetic ordered systems can be calculated by~\cite{Lovesey,SpinW}
\begin{equation}
I\left(\mathbf{Q},\hbar\omega\right)\propto\sum_{\alpha,\beta}^{x,y,z}\left(\delta_{\alpha,\beta}-Q^\alpha Q^\beta/Q^2\right)S^{\alpha\beta}\left(\mathbf{Q},\hbar\omega\right),\label{Eq:SWIntensity}
\end{equation}
where $S^{\alpha\beta}\left(\mathbf{Q},\hbar\omega\right)$ is a dynamic structure factor calculated from the eigenstates of Eq.~(\ref{Eq:Hamiltonian}), the Bose factor $[1-\text{exp}(-\hbar\omega/k_\text{B}T)]^{-1}$, and the magnetic form factor $f(Q)$.
The obtained $I\left(\mathbf{Q},\hbar\omega\right)$ was then powder averaged and convoluted with the instrumental energy resolution function~\cite{HRC_1,4SEASONS_4} for comparison with the experimental results.
The optimum parameters in Eq.~(\ref{Eq:Hamiltonian}) were determined by fitting calculated to experimental $I(\hbar\omega)$ at $Q\sim1.2\ ~\text{\AA}^{-1}$ [Fig.~\ref{Fig:Cuts}(d)], $I(Q)$ at $\hbar\omega\sim11.5\ ~\text{meV}$ [Fig.~\ref{Fig:Cuts}(e)], $I(Q)$ at $\hbar\omega\sim21.5\ ~\text{meV}$ [Fig.~\ref{Fig:Cuts}(f)], and the $Q$ dependences of the peak positions [Fig.~\ref{Fig:Map}(c)].
We fit all spectra simultaneously to reduce the ambiguity of obtained parameters.
We started with the parameters $J_1=8.62$~meV and $J'=0.01$~meV previously obtained by single-crystal bulk magnetization measurements.~\cite{La5Mo4O16_Konishi}
The obtained optimum parameters are 
\begin{eqnarray}
J_1&=&6.06(2)~\text{meV},\nonumber\\
\delta_1&=&0.80(1),\nonumber\\
J_2&=&0.78(1)~\text{meV},\label{Eq:Parameters}\\
\delta_2&=&0.94(3),\nonumber\\
J'&=&0.035(8)~\text{meV}, \nonumber\\
D&=&-1.71(4)~\text{meV}.\nonumber
\end{eqnarray}
The fitting results are described by solid lines in Figs.~\ref{Fig:Map}(c) and \ref{Fig:Cuts}(d)--\ref{Fig:Cuts}(f).
Good agreements between the experiments and LSW calculations can be seen.
Note that we only used one overall scaling factor to calculate the neutron scattering intensities in Figs.~\ref{Fig:Cuts}(d)--\ref{Fig:Cuts}(f).
Furthermore, the calculated $I(Q,\hbar\omega)$ map at 4~K using the optimum parameters is also shown in Fig.~\ref{Fig:Map}(b).
Good agreements can be seen again.

The observed spin gap [Fig.~\ref{Fig:Map}(d)] is well explained by the LSW analysis; the calculated spin-wave energies at one of the magnetic zone centers $\mathbf{Q}=(1,0,0.5)$ using the optimum parameters in Eq.~(\ref{Eq:Parameters}) are $9.56$, $9.91$, $18.67$, and $21.25$~meV.
The easy-axis anisotropy ($\delta_1$, $\delta_2<1$ and $D<0$) and very weak interlayer exchange interaction ($J'/J_1<0.01$) in the obtained model Hamiltonian support the microscopic models proposed for the origin of the high-temperature magnetoresistance~\cite{La5Mo4O16_Kobayashi} and long-time magnetization decay in La$_5$Mo$_4$O$_{16}$.~\cite{La5Mo4O16_Konishi}
As reported above, the ordered magnetic moments at Mo1 and Mo2 are reduced to about half of the spin-only values.
Since the valences of the Mo1 and Mo2 sites are about +5 and +4, respectively, at all measured temperatures, as shown in Fig.~\ref{Fig:Diffraction}(d), anomalous charge disproportionation of the Mo $4d$ electrons between Mo1 and Mo2 sites in the long-range magnetic ordered state is ruled out.
Instead, our INS study suggests that fluctuations induced by low dimensionality and $J_2$ bond frustration [see $J'/J_1\ll1$ and $J_2>0$ in Eq.~(\ref{Eq:Parameters})] are the origin of the reduced magnetic moments.
The quantum fluctuation effect due to the spin $S=1/2$ (and 1) may also be the origin of the reduced moments.
The small magnetic moments at Mo1 and Mo2 and the large spin gap suggest that the SOI in La$_5$Mo$_4$O$_{16}$ induces the strong easy-axis magnetic anisotropy instead of the total momentum number $\mathbf{J}$.
Therefore, our results indicate that the SOI in La$_5$Mo$_4$O$_{16}$ is similar to that in $3d$ electron systems rather than that in $5d$ and $4f$ electron systems.
In fact, similar spin gaps in the long-range magnetic ordered state were observed in $3d$~\cite{Ca3Co2O6} and other $4d$ electron systems.~\cite{Ba2YRuO6,La2NaRuO6,Ca2RuO4}

\section{Summary}
The magnetic structure and spin dynamics in the localized $4d$ electron layered perovskite antiferromagnet La$_5$Mo$_4$O$_{16}$ have been studied by TOF powder neutron measurements.
Our NPD refinement revealed a magnetic structure where Mo1 and Mo2 are coupled antiferromagnetically within the layers and the layers are also coupled antiferromagnetically along the $c$ axis with reduced magnetic moments.
INS results were analyzed using a model Hamiltonian including the easy-axis anisotropy and weak interlayer exchange interaction using semiclassical LSW theory.
The spin gap at the magnetic zone center is explained by Ising anisotropy due to the intermediate scale of the SOI.
The microscopic spin structure, strong easy-axis anisotropy, and very weak interlayer exchange interaction compared with the intralayer exchange interactions determined by our TOF elastic and inelastic neutron scattering measurements support the microscopic models proposed for the high-temperature magnetoresistance and long-time magnetization decay in La$_5$Mo$_4$O$_{16}$.

\section*{Acknowledgements}
We thank Y. Ishikawa and T. Kamiyama for supporting the Rietveld refinement, and S. Yano, C. M. Wu, and J. S. Gardner for preliminary neutron diffraction measurements.
We also thank M. Sato, T. J Sato, and A. Nakao for helpful discussion.
The proposal numbers of the experiments at 4SEASONS, iMATERIA, and HRC in MLF, J-PARC are 2014A0082, 2014A0258, and 2014B0141, respectively.
This work was partly supported by JSPS KAKENHI Grant Numbers JP15K04742, JP17K14349, and JP25287090, and the Cooperative Research Program of ``Network Joint Research Center for Materials and Devices'' (2015143 and 20161060).


\end{document}